\documentclass[aps,prl,twocolumn,superscriptaddress,amssymb,nofootinbib]{revtex4}

\usepackage{color}
\usepackage{epsfig}
\usepackage{graphicx}
\usepackage{epstopdf}

\begin{document}

%\preprint{UCD-HEP-???}
\title{Light Charged Higgs and Lepton Universality in W boson Decays}
%\title{Light Charged and CP-odd Higgses in Supersymmetric Models}

\author{Radovan Derm\' \i\v sek}
\email[]{dermisek@ias.edu}

%\homepage[]{Your web page}
        %\thanks{}
%\altaffiliation {}
\affiliation{School of Natural Sciences, Institute for Advanced Study, Princeton,
NJ 08540}

%\author{}
%\email[]{}

%\affiliation{}

\date{\today}
%\date{January 5, 2006}

\begin{abstract}
We study the effect of a light charged Higgs appearing in supersymmetric models containing two Higgs doublets on the measurement of leptonic branching ratios of the W boson at LEP. We show that the $2.8 \sigma$ excess of the branching ratio $W \to \tau \nu$ with respect to the other leptons correlates well with the existence of charged Higgs with mass close to the mass of the W boson which dominantly decays into $W^\star$ and a light CP odd Higgs boson $A$ with mass below $2m_b$, so that it decays into $\tau^+ \tau^-$ and $c \bar c$. There are no searches for the charged Higgs in this channel and thus it could be discovered in LEP data
or at the Tevatron where it would be frequently produced in top quark decays.

\end{abstract}

% insert suggested PACS numbers in braces on next line
\pacs{}
% insert suggested keywords - APS authors don't need to do this
\keywords{}

%\maketitle must follow title, authors, abstract, \pacs, and \keywords
\maketitle

% body of paper here - Use proper section commands
% References should be done using the \cite, \ref, and \label commands
%\section{ \label{sec:}}
% Put \label in argument of \section for cross-referencing
%\section{\label{}}

%\subsection{}

%\subsubsection{}

%\subsection{Introduction}

%\subsubsection{Experimental results}

%%%%%%%%%%%%%%%%%%%%%%%%%%%%%%%%%%%%%%%%%%%%%%%%
%\subsection{Introduction}
%%%%%%%%%%%%%%%%%%%%%%%%%%%%%%%%%%%%%%%%%%%%%%%%

{\it Introduction:}
The existence of a pair of charged Higgs bosons is predicted by several extensions of the standard model.
For example, in the minimal supersymmetric standard model (MSSM) the Higgs sector contains two Higgs doublets which lead to five Higgs bosons in the spectrum:  light and heavy CP even Higgses, $h$ and $H$, the CP odd Higgs, $A$, and a pair of charged Higgs bosons, $H^\pm$. The discovery of Higgs bosons is an important step in understanding electroweak symmetry breaking and uncovering the ultimate theory of particle physics.

The presence of the charged Higgs can manifest itself in various ways. Charged Higgs contributes through quantum corrections to all electroweak observables or it can be directly produced in $e^+ e-$ collisions or it can show up in decay products of heavier particles, e.g. the top quark. While quantum corrections  to electroweak observables can be canceled by contributions of other particles in a given model, the signs of a direct production of charged Higgs cannot be erased by additional particles. In this letter we show that the measured discrepancy in lepton universality in $W$ boson decays can be interpreted as a direct production of the charged Higgs with mass close to the mass of the $W$ boson in MSSM-like models.

%%%%%%%%%%%%%%%%%%%%%%%%%%%%%%%%%%%%%%%%%%%%%%%%
%\subsection{$W \to  \tau \nu$ at LEP and Tevatron}
%%%%%%%%%%%%%%%%%%%%%%%%%%%%%%%%%%%%%%%%%%%%%%%%

{\it $W \to  \tau \nu$ at LEP and the Tevatron:}
From the combined results of LEP collaborations on the 
leptonic branching ratios of the W boson an excess of the branching ratio 
$W \to \tau \nu$ with respect to the other leptons is evident~\cite{:2004qh}.
While measured branching ratios $W \to e \nu$ and $W \to \mu \nu$ perfectly agree with 
lepton universality,
\begin{equation}
B(W \to \mu \nu) / B(W \to e \nu) = 0.994 \pm 0.020,
\end{equation}
the branching fractions in taus with respect to electrons and muons differ by more than two standard deviations:
\begin{eqnarray}
B(W \to \tau \nu) / B(W \to e \nu) &=& 1.070 \pm 0.029,\\
B(W \to \tau \nu) / B(W \to \mu \nu) &=& 1.076 \pm 0.028.
\end{eqnarray}
The ratio between the tau fraction and the average of electron and muon fractions
\begin{equation}
R_{\tau/l} \equiv 2B(W \to \tau \nu) / (B(W \to e \nu)+B(W \to \mu \nu)) ,
\end{equation}
\begin{equation}
R_{\tau/l}^{exp} = 1.073 \pm 0.026,
\label{eq:Rtau_l_exp}
\end{equation}
results in a poor agreement, at the level of 2.8 standard deviation, with lepton universality predicted by the standard model (SM).

The $WW$ pair production cross section, $\sigma_{W^+W^-}$, at LEP is about 17 pb at the center of mass energy $\sqrt{s} = 200$ GeV and $W^\pm$ decay equally (in the SM) to each generation of leptons with branching ratio of $10.6 \%$. Couplings of charged Higgs to fermions are proportional to the mass of the charged fermion and thus the charged Higgs preferably decays into $\tau \nu$ while decays to first two generations of leptons are highly suppressed by ratios of fermion masses squared $m_\mu^2 / m_\tau^2 \simeq 0.003$ and $m_e^2 / m_\tau^2 \simeq 8 \times 10^{-8}$. Since charged Higgs pair production cross section, $\sigma_{H^+H^-}$, is about 160 fb for $m_{H^\pm} \simeq m_{W^\pm}$, about two orders of magnitude smaller than $\sigma_{W^+W^-}$, and charged Higgs may decay to $\tau \nu$ with significantly larger branching fraction than $W$ (depending on the parameter space) already a naive estimate suggests that a charged Higgs with mass close to the mass of the $W$ boson can easily contribute to the measurement of lepton universality at LEP at the level indicated by the experimental result~(\ref{eq:Rtau_l_exp}).

Lepton universality in $W$ decays was measured also at 
the Tevatron~\cite{Safonov:2004zv, Rimondi:1999pu}. 
CDF~\cite{Safonov:2004zv} is looking at inclusive W production
%, $\sigma (p \bar p \to W)*Br(W \to \tau \nu) \simeq 2.62$ nb, 
and the ratio
 $Br(W \to \tau \nu)/Br(W \to e \nu)  = 0.99 \pm 0.04(stat) \pm 0.07(syst)$
agrees with lepton universality.
W bosons are produced in $p \bar p$ interactions dominantly through the Drell-Yan process, e.g.
$u \bar d \to W$, for which the production cross section is $\sigma (p \bar p \to W)*Br(W \to \tau \nu) 
\simeq 2.62$ nb. The production cross section of a single charged Higgs from first-generation quarks is obviously negligible. In addition, the $WW$ pair production cross section is also negligible, $\sim 15$ pb, 
 and thus a single (or a pair) production of charged Higgs(es) is not expected to affect lepton universality 
in this measurement.

Direct production of the charged Higgs boson with mass close to the mass of W boson is a unique way to explain the deviation from lepton universality in W decays at LEP and agreement with lepton universality in W decays measured at the Tevatron. Any possible alternative explanation by new physics that would modify the $W\tau \nu$ vertex through loop corrections would necessarily predict the deviation from lepton universality at both LEP and the Tevatron.

%%%%%%%%%%%%%%%%%%%%%%%%%%%%%%%%%%%%%%%%%%%%%%%%
%\subsection{$m_A \ll m_W$ and $\tan \beta \simeq 1$ scenario}
%%%%%%%%%%%%%%%%%%%%%%%%%%%%%%%%%%%%%%%%%%%%%%%%

{\it The $m_A \ll m_W$ and} $\tan \beta \lesssim 2.5$ {\it scenario:}
In the MSSM the mass of the charged Higgs is given as,
\begin{eqnarray}
m_{H^\pm} &=& \sqrt{m_W^2 + m_A^2 - \Delta^\prime} \; , 
\end{eqnarray}
where $m_A$ is the mass of the CP odd Higgs boson and $\Delta^\prime$ represents radiative correction from superpartners which is typically not significant  (it is positive and has a tendency to decrease the mass of the charged Higgs).
Thus, $m_{H^\pm} \simeq m_W$ requires $m_A \ll m_W$. 

Although this scenario is ruled out in the MSSM (only by searches for the CP even Higgs boson), it has been recently argued that for $m_A < 2m_b$ and $\tan \beta \lesssim 2.5$  the scenario is the least constrained and thus easily viable in simple extensions of the MSSM~\cite{Dermisek:2008id}. The reason is that for $m_A \ll m_W$ and $\tan \beta \simeq 1$ the light CP even Higgs boson becomes SM-like, 
%$C_{ZZh} \simeq 1$, since $C_{ZZH} \simeq 0$, see Eq.~(\ref{eq:ZZH_mAlessmZ}), 
and although
it is massless at the tree level, it will receive a contribution from superpartners and the tree level relation between the light CP even and CP odd Higgses, $m_h < m_A$, is dramatically changed by SUSY corrections. Even for modest superpartner masses the light CP even Higgs boson will be heavier than $2m_A$, for superpartner masses between 300 GeV and 1 TeV we find $m_h \simeq 40 - 60 $ GeV,  and thus $h \to AA$ decay mode is open and generically dominant. 

Since $h$ is SM-like,  $e^+ e^- \to hA$ is highly suppressed and the limits from the Z width measurements can be easily satisfied even for  $m_h + m_A < m_Z$. 
On the other hand, $h$ would be produced in association with the Z boson. However, for small $\tan \beta$ the width of $A$ is shared between $\tau^+ \tau^-$ and $c \bar c$ for $m_A < 2m_b$ and thus the width of $h$ is spread over several different final states, $4\tau$, $4c$, $2\tau 2c$ and highly suppressed $b \bar b$ and thus the LEP limits in each channel separately are highly weakened. However the decay mode independent limit requires the SM like Higgs to be above 82 GeV which rules this scenario out in the MSSM,
since $m_h$ cannot be pushed above 82 GeV by radiative corrections.

The rest of the Higgs spectrum is basically not constrained at all in this scenario. 
The heavy CP even and the CP odd Higgses could have been produced at LEP in $e^+ e^- \to H A$ but they would avoid detection because $H$ dominantly decays to $ZA$ - the mode  that has not been searched for. Additional constraints are discussed in detail in Ref.~\cite{Dermisek:2008id}.
The charged Higgs is also very little constrained as we discuss later.

The mass of the light CP even Higgs is the only problematic part in this scenario.
There are however various ways to increase the mass of the SM-like Higgs boson in extensions of the MSSM. A simple
possibility is to consider singlet extensions of the MSSM containing $\lambda S H_u H_d$ term in the superpotential.
It is known that this term itself contributes $\lambda^2 v^2 sin^2 2 \beta$, where v = 174 GeV, to the mass squared of
the CP even Higgs~\cite{Ellis:1988er} and thus can easily push the Higgs mass above the decay-mode independent limit, 
$82$ GeV.\footnote{Singlet extensions can also alter the couplings of the Higgses to $Z$ and
$W$ through mixing~\cite{mixed} or provide new Higgs decay modes~\cite{4tau, Chang:2005ht, Chang:2008cw}. 
We do not consider these
possibilities since they would lead to model dependent
predictions.}
 Note, this contribution
is maximized for $\tan \beta \simeq 1$. 
In this paper we assume that a possible extension does not significantly alter the two Higgs
doublet part of the Higgs sector besides increasing the Higgs mass above the decay-mode independent limit.
Thus in the discussion of the charged Higgs contribution to the measurement of lepton universality at LEP we 
use exact MSSM couplings and branching ratios of the charged Higgs.\footnote{ 
This is not an unreasonable assumption, it is usually the case that an extension of a given model has a limit in which it resembles the original model, e.g. the MSSM in the decoupling limit resembles the standard model, the next-to-minimal supersymmetric model (NMSSM) has a limit in which it resembles the MSSM and so on. Indeed, in the NMSSM the scenario with a light MSSM like CP odd Higgs and small $\tan \beta$ is viable and has all the features of the MSSM in this limit~\cite{NMSSM_small_tb}. It should be stressed however that this scenario is not limited to singlet extensions of the MSSM and it would be viable in many models beyond the MSSM that increase the mass of the SM-like Higgs boson.}
In the MSSM for $m_A < 2m_b$ and $1 \lesssim \tan \beta \lesssim 2.5$ we find $m_{H^\pm} \simeq m_W$ and varying
$\tan \beta$  between 1 and $2.5$ we have~\cite{Dermisek:2008id}:
\begin{equation}
B (H^+ \to W^{+ \star } A, \; \tau^+ \nu)  \; \simeq \;  70 \%, \; 20 \% \;  - \;  35 \%, \; 65 \%,
\label{eq:BHpm}
\end{equation}
with $B (H^+ \to c \bar s) \simeq 10 \%$ for $\tan \beta = 1$ which becomes negligible for $\tan \beta \gtrsim 1.5$.
For the discussion of experimental constraints let us also include branching ratios of the top quark,
\begin{equation}
B(t \to H^+ b, \; W^+ b) \; \simeq \; 40 \%, \; 60 \% \; - \; 10 \%, \; 90 \% ,
\label{eq:Bt}
\end{equation}
again varying $\tan \beta$ in $ 1 - 2.5$ range. These results are not very sensitive to superpartner masses nor the mass of the CP odd Higgs as far as $m_A < 2m_b$.

%%%%%%%%%%%%%%%%%%%%%%%%%%%%%%%%%%%%%%%%%%%%%%%%
%\subsection{Charged Higgs at LEP and Tevatron}
%%%%%%%%%%%%%%%%%%%%%%%%%%%%%%%%%%%%%%%%%%%%%%%%

{\it Experimental limits on charged Higgs:}
A search for pair produced charged Higgs bosons was performed by LEP 
collaborations~\cite{:2001xy, Achard:2003gt, Heister:2002ev, Abdallah:2003wd}.
A pair of charged Higgs bosons can be produced in $e^+ e^-$ collisions via s-channel exchange of 
a $Z$-boson or a photon. 
Three different final states, $\tau^+ \nu \tau^- \bar \nu$, $c \bar s \bar c s$ and
$c \bar s \tau^- \bar \nu$ were considered
and lower limits were set on the mass $m_{H^\pm}$ 
as a function of the branching ratio $B(H^+ \to \tau^+ \nu)$, assuming 
$B(H^+ \to \tau^+ \nu) + B(H^+ \to c \bar s)=1$.
In addition, DELPHI considered a possibility $H^+ \to W^{+\star} A$ which is important if the CP 
odd Higgs boson is not too heavy~\cite{Akeroyd:2002hh} and limits were obtained under the assumption that the pseudoscalar is heavy
enough to decay into $b \bar b$~\cite{Abdallah:2003wd}.

The topology  of the $H^+ H^-$ signal is very similar to the $W^+ W^-$ pair production which is the dominant background. 
Pairs of $W^\pm$ are produced in $e^+ e^-$ collisions via s-channel exchange of 
a $Z$-boson or a photon in addition to t-channel exchange of a neutrino.
To discriminate charged Higgs from W boson events jet flavor tagging (c/s) and the difference in polarization of taus originating from $W^\pm$ and $H^\pm$ are used in some analyzes.

The strongest limits are set by ALEPH~\cite{Heister:2002ev}. Assuming $B(H^+ \to \tau^+ \nu) + B(H^+ \to c \bar s)=1$, charged Higgs bosons with mass below 79.3 GeV are excluded at 95\% C.L., independent of $B(H^+ \to \tau^+ \nu)$. Somewhat lower limits have been obtained by DELPHI~\cite{Abdallah:2003wd} and L3~\cite{Achard:2003gt} collaborations due to local excesses of events.

In the scenario discussed above the charged Higgs can decay dominantly into $W^\star A$ with $A \to c \bar c$ or $\tau^+ \tau^-$~(\ref{eq:BHpm}). LEP limits thus apply to the remaining branching ratios and are comfortably satisfied for  $m_{H^\pm} \gtrsim 75$ GeV.

At the Tevatron the charged Higgs is searched for in the decay of the top quark~\cite{Abulencia:2005jd, Affolder:1999au, Abe:1997rk, Abazov:2001md, Abbott:1999eca}. The production of $t \bar t$ pairs with a cross section of 6.7 pb could be a significant source of charged Higgses. If kinematically allowed, the top quark can decay to $H^+ b$, competing with the standard model decay $W^+ b$. The strongest limits come from CDF~\cite{Abulencia:2005jd} which uses measurements of the $t \bar t$ production cross section in the $l+ \slash \hspace{-.25cm} E_T + jets + X$ channels, where $l=e, \mu$ and $X=l, \tau$ or tagged jets from
data samples corresponding to  an integrated luminosity of 193 ${\rm pb}^{-1}$.
It is assumed that the charged Higgs can decay only to $\tau^+ \nu$, $c\bar s$, $t^\star \bar b$ or $W^+ A$ with $A \to b \bar b$.

If charged Higgs decays exclusively to $\tau^+ \nu$, the $B(t \to H^+ b)$ is constrained to be less than 0.4 at 95 \% C.L. 
For MSSM benchmark scenarios, assuming $H^+ \to \tau^+ \nu$ or $H^+ \to c \bar s$ only, stronger limits than at LEP are set for $\tan \beta \lesssim 1.3$ on the mass of the charged Higgs. For  $\tan \beta \lesssim 1$ the limit is $m_{H^\pm} \gtrsim 100$ GeV. 
If no assumption is made on the charged Higgs decay (but still allowing only those that were searched for) the  $B(t \to H^+ b)$ is constrained to be less than $\sim 0.8$ for $m_{H^\pm} \simeq 80$ GeV.

If charged Higgs decays dominantly into $W^\star A$ with $A \to c \bar c$ or $\tau^+ \tau^-$,
the decay modes that were not search for, and in addition modes that can easily mimic $W$ decay modes, especially the dominant hadronic mode, it is reasonable to expect that the limits would be somewhat weaker. Since in our scenario $B(t \to H^+ b) \lesssim 40 \%$~(\ref{eq:Bt}),  the Tevatron  does not place stronger limits than LEP. 

For small $\tan \beta$ charged Higgs with mass close to the mass of $W$ contributes negligibly to the $W \tau \nu$ vertex
and it also does not significantly modify $Z b_{L,R} \bar b_{L,R}$ vertices. In addition, these contributions are comparable with possible contributions from superpartners. However, the contribution of a light charged Higgs to $b \to s \gamma$ is sizable and has to be canceled by the chargino-stop contribution which can be of the same size or larger with an opposite sign for light chargino and stop and large mixing in the stop sector. A light charged Higgs would also contribute to $B \to \tau \nu$ at the tree level. Its contribution scales as $\tan^2 \beta$ and for small $\tan \beta$ it is well withing experimental limits~\cite{Nierste:2008qe}.

%%%%%%%%%%%%%%%%%%%%%%%%%%%%%%%%%%%%%%%%%%%%%%%%
%\subsection{Charged Higgs contribution  to lepton non-universality}
%%%%%%%%%%%%%%%%%%%%%%%%%%%%%%%%%%%%%%%%%%%%%%%%

\begin{figure}[t]
\includegraphics[width=3.2in]{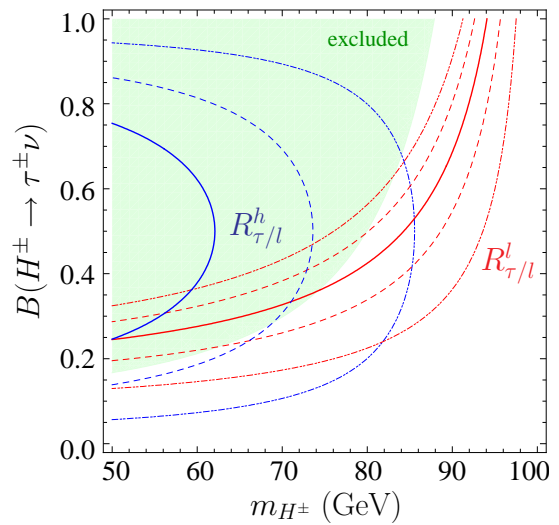}
\caption{$R_{\tau/l}^l$ (red) and $R_{\tau/l}^h$ (blue) as a function of $m_{H^\pm}$ and $B(H^\pm \to \tau \nu)$ for $\sqrt{s} = 200$ GeV. Solid lines represent $R_{\tau/l}^l, R_{\tau/l}^h = R_{\tau/l}^{exp} = 1.073$ and dashed and dotted lines indicate $1\sigma = \pm 0.026$ and $2\sigma$ ranges. Shaded region is excluded by LEP searches for the charged Higgs boson, assuming $B(H^\pm \to \tau \nu) = 1$. Other limits apply for $m_{H^\pm} \lesssim 75$ GeV that are not easy to implement in the plot (see the text).}
\label{fig:RWtaunu}
\end{figure}

{\it Charged Higgs contribution  to lepton non-universality:}
Charged Higgs can contribute in the fully leptonic
$\tau \nu \tau \nu$ and semi-leptonic $\tau \nu + hadrons$ 4-fermion production channels.
Its contribution in the $\tau \nu \tau \nu$ channel would manifest itself in the excess of 
$\tau \nu \tau \nu$ events compared to $l \nu l \nu$, $l = e,\mu$ events and 
would be attributed to the
larger branching ratio of $W \to \tau \nu$ compared to $W \to l \nu$,  $l = e,\mu$.
This increase is given by
\begin{equation}
R_{\tau/l}^l = \sqrt{1 + \frac{\sigma_{H^+ H^-} B(H^+ \to \tau^+ \nu)^2}
{\sigma_{W^+ W^-} B(W^+ \to l^+ \nu)^2}}.
\end{equation}
Charged Higgs can contribute directly only to $\tau \nu \tau \nu$ channel and not to mixed 
$\tau \nu l \nu$,  $l = e,\mu$ channels. However if $\tau$ decays leptonically the 
efficiency of an  $W \to \tau \nu$ event to pass as a $W \to l \nu$ event
 is not small and so the charged Higgs production would effectively contribute 
to both $\tau \nu \tau \nu$ and mixed $\tau \nu l \nu$ channels. 
For this reason the prediction of $R_{\tau/l}^l$ should be treated only as an estimate of the
effect of the charged Higgs on lepton non-universality in $W$ decays. 

In a similar way the contribution to the $\tau \nu + {\rm hadrons}$ final state that 
would be attributed to the
larger branching ratio of $W \to \tau \nu$ compared to $W \to l \nu$,  $l = e,\mu$
can be roughly estimated by
\begin{equation}
R_{\tau/l}^h = 1 + \frac{\sigma_{H^+ H^-} B(H^+ \to \tau^+ \nu) B(H^+ \to hadrons)}
{\sigma_{W^+ W^-} B(W^+ \to l^+ \nu)B(W^+ \to hadrons)}
\label{eq:Rtaulhad}
\end{equation}
with
\begin{equation}
B(H^+ \to hadrons) \simeq 1- B(H^+ \to \tau^+ \nu).
\label{eq:BHtohadrons}
\end{equation}
In this case the situation is not so simple as for the fully leptonic channel and the above formula should be considered as a rough estimate of the effect the charged Higgs would have on the lepton non-universality in $W$ decays. On one hand the formula above overestimates the hadronic branching 
ratio since $1- B(H^+ \to \tau^+ \nu) = B(H^+ \to c \bar s) + B(H^+ \to W^{+\star} A)$ and the part of $B(H^+ \to W^{+\star} A)$ for which $A \to \tau^+ \tau^-$ and $W^{+\star} \to leptons$ should not be counted in $B(H^+ \to hadrons)$, although if at least two taus from $A$ or $W$ decay hadronically it still might be counted as hadronic decay of $H^+$. On the other hand the formula does not take into account 
events resulting from the dominant decay mode of the charged Higgs, $W^\star A$, of the type: $H^+ H^- \to c \bar s W^{-\star} A$, $\bar c s W^{+\star} A$, $ W^{+\star} A W^{-\star} A$ in which one of the $A \to \tau^+ \tau^-$ that could mimic $WW \to \tau + hadrons$. 
To estimate the efficiency for these events to pass the selection cuts for WW production would require a careful analysis of LEP collaborations. Although it might be a significant contribution to the lepton non-universality we neglect this contribution and will treat $R_{\tau/l}^h$ given in Eq.~(\ref{eq:Rtaulhad}) as a rough approximation (quite likely an underestimation) of the effect of the charged Higgs on the lepton 
non-universality measured in $W$ decays.

In Fig.~\ref{fig:RWtaunu} we show $R_{\tau/l}^l$ (red) and $R_{\tau/l}^h$ (blue) as a function of $m_{H^\pm}$ and $B(H^\pm \to \tau \nu)$ for $\sqrt{s} = 200$ GeV. Solid lines represent $R_{\tau/l}^l, R_{\tau/l}^h = R_{\tau/l}^{exp} = 1.073$ and dashed and dotted lines indicate $1\sigma = \pm 0.026$ and $2\sigma$ ranges. Shaded region is excluded by LEP searches for the charged Higgs boson, assuming $B(H^\pm \to \tau \nu) = 1$. Other limits apply for $m_{H^\pm} \lesssim 75$ GeV  as we discussed before 
but these are not easy to implement in the plot because they depend on other parameters, e.g. $\tan \beta$. We see that the charged Higgs with mass $75 - 85$ GeV and $B(H^+ \to \tau^+ \nu) \simeq 20 - 60 \%$ has the right properties to explain the measured deviation from lepton universality in $W$ decays. The properties of the charged Higgs favored by the $R_{\tau/l}^{exp} $ are exactly those found in the  $m_A \ll m_W$, $\tan \beta \lesssim 2.5$ scenario~(\ref{eq:BHpm}).

Clearly the search for the charged Higgs including the dominant  $W^\star A$ with $A \to c \bar c$ or $\tau^+ \tau^-$ decay modes at LEP and especially at the Tevatron with currently available much larger data sample is very desirable.

    {\bf Note added:} after completion of this work we became aware of the work of J. H. Park~\cite{Park:2006gk}
where the possibility of a charged Higgs explanation of the lepton non-universality in $W$ boson decays was discussed. 
To explain the lepton non-universality and avoid experimental constraints a general two Higgs doublet model 
was considered. 
The mass of the charged Higgs and its couplings to fermions needed to explain the non-universality and to avoid other experimental constraints are freely adjustable parameters. This scenario does not have a supersymmetric extension. The possibility we present might be more compelling since we consider the MSSM-like charged Higgs for which couplings and mass relations to other Higgses are fixed. It is the small $m_A$ that plays a multiple role here. First of all it make the whole scenario easily phenomenologically viable in simple extensions of the MSSM, it leads to $m_{H^\pm} \simeq m_{W^\pm}$ and it is also responsible for reduced $B(H^+ \to \tau^+ \nu)$
as a result of dominant  $B(H^+ \to W^{+\star} A)$ that is sufficient to avoid experimental limits from LEP and the Tevatron searches and explain the lepton non-universality in $W$ decays.

%%%%%%%%%%%%%%%%%%%%%%%%%%%%%%%%%%%%%%%%%%%%%%%%
\acknowledgments
%%%%%%%%%%%%%%%%%%%%%%%%%%%%%%%%%%%%%%%%%%%%%%%%

I would like to thank J. Gunion, B. McElrath, S. Mishima, M. Papucci, J. H. Park and M.~A.~Sanchis-Lozano
for useful discussions.
This work is supported by the U.S. Department of Energy, grant DE-FG02-90ER40542.  

\vspace{0.2cm}

%%%%%%%%%%%%%%%%%%%%%%%%%%%%%%%%%%%%%%%%%%%%%%%%

%%%%%%%%%%%%%%%%%%%%%%%%%%%%%%%%%%%%%%%%%%%%%%%%

%%%%%%%%%%%%%%%%%%%%%%%%%%%%%%%%%%%%%%%%%%%%%%%%
\end{document}